\documentstyle[preprint,aps,prb]{revtex}
\begin{document}
\draft
\title{\bf Transport properties of a quantum dot with superconducting leads}
\author{C. B. Whan and T. P. Orlando} 
\address{Department of Electrical Engineering and Computer Science,\\
Massachusetts Institute of Technology, Cambridge, MA 02139}
\date{\today}
\maketitle
\begin{abstract}
We report a numerical study of transport properties of a 
quantum dot with superconducting leads. We introduce a 
general phenomenological model of quantum dot transport, 
in which electron tunnel rates are computed within the 
Fermi's Golden Rule approach. The low temperature current-voltage 
(I-V) characteristics are in qualitative agreement with 
experimental observations of Ralph {\it et al.} {[Phys. Rev. Lett., 
{\bf 74}, 3241 (1995)]}. At higher temperatures, our results 
reveal new effects due to the thermal excitation of quasiparticles 
in the leads as well as the thermal population of excited quantum 
levels in the dot. We also study the photon-assisted tunneling 
phenomena in our system and point out its potential for millimeter 
wave applications.
\end{abstract}

Due to the advancement of modern lithographic technologies, electrons 
can now be confined to a spatial region so small that not only the 
Coulomb charging energy $E_C$, but also the spacing of the quantum 
mechanical energy levels $\epsilon$, become accessible to experiments 
conducted at low temperatures. While the single electron 
charging effect has been observed in both metal tunnel junction 
systems\cite{fulton} and semiconductor quantum dots,\cite{scott-thomas} 
the discrete quantum levels have been studied mostly in semiconductor 
quantum dots.\cite{mcuen,johnson} Recently, Ralph, Black and Tinkham 
conducted experiments in which they directly measured tunneling 
transport through discrete quantum levels in a small metallic 
particle.\cite{ralph}

In this paper we present numerical simulation results based on the 
system studied by Ralph {\it et al.}\cite{ralph} Their samples were made 
of a tiny aluminum particle (with diameter $<$ 10 nm) connected 
to aluminum leads through tunnel junctions on both sides. A third 
electrode can also be added to act as a gate. Since the experiment 
was conducted at temperatures well below the superconducting transition 
temperature of aluminum ($T_C = 1.21\,$K for their sample), the leads 
are nominally superconducting unless a magnetic field is applied. 
The dot itself can have discrete quantum levels with spacing 
either smaller or larger than the superconducting energy gap 
of aluminum.\cite{ralph,black,anderson} In this paper, we will 
assume that the quantum dot is a normal metal particle, and the 
leads can be either superconducting or normal. 

We introduce a general phenomenological model of transport 
through a quantum dot. In this model, we assume that the quantum dot 
is weakly coupled to the two leads by tunnel barriers. When an appropriate  
bias voltage $V$ is applied to the leads, an electron can tunnel across  
one barrier into the dot and subsequently tunnel out through the second 
barrier. According to general tunneling theory,\cite{solymar,averin} 
the tunneling rate across a barrier from side ``a'' to side ``b'', can 
be evaluated using the Fermi's Golden Rule,
\begin{eqnarray}
\label{eqn:golden-rule}
\Gamma_{a \rightarrow b} (\mu_{a}, \mu_{b}) &= \frac{2\pi}{\hbar} \int_{-\infty}^{
\infty} 
{|T_{ab}|}^{2} {\mathcal{N}}_{a}(E - \mu_{a}) 
{\mathcal{N}}_{b}(E - \mu_{b}) \nonumber \\ 
&\times f(E - \mu_{a}) \left[ 1 - f(E - \mu_{b}) \right] dE
 , 
\end{eqnarray}
where $T_{ab}$ is the phenomenological tunneling matrix element, and  
$f(x) = 1/[1 + \exp(x/k_{B}T)]$ is the Fermi function.\cite{note0,beenakker} 
${\mathcal{N}}_{a}(E)$ and ${\mathcal{N}}_{b}(E)$ are the density of states, 
and $\mu_a$ and $\mu_b$ are the chemical potentials, on their corresponding 
sides. For our system, to compute the tunneling rate from one of 
the leads to the dot, we take the BCS quasiparticle density of 
states\cite{tinkham} in the lead and assume that the dot itself has an 
evenly spaced (with spacing $\epsilon$) discrete level spectrum. Thus the 
density of states in the superconducting leads is,
\begin{equation}
\label{eqn:Sdos}
{\mathcal{N}}_{S}(E) = {\mathcal{N}}_{N} \left\{ 
\begin{array}{ll} 
0 & \mbox{if $|E| \le \Delta$}; \\
\frac{| E |}{\sqrt{E^2 - {\Delta}^2}} & \mbox{if $|E| > \Delta$},
\end{array} 
\right. 
\end{equation}
and the density of states in the dot is,
\begin{equation}
\label{eqn:Qdos}
{\mathcal{N}}_{D}(E) = {\mathcal{N}}_{N} \, \epsilon \, \sum_{n}{\delta (E - n \epsilon)}. 
\end{equation}
Here, $2\Delta$ is the superconducting energy gap of the leads, and 
${\mathcal{N}}_{N}$ is the density of states in the bulk normal metal 
that comprises both the leads and the dot. 

Once we have the tunneling rates, we can compute the tunneling current 
through the dot using a master equation approach.\cite{averin} At steady 
state, the current that flows into the dot from one lead should balance the 
current that flows out of the dot into the other lead. Therefore the 
steady state current is given by, 
\begin{equation}
\label{eqn:IV}
I = \sum_{k} \left[ \Gamma_{L \rightarrow D}(\mu_{L}, \mu_{k}) P_{k-
1} - \Gamma_{D \rightarrow L}(\mu_{k}, \mu_{L}) P_{k} \right], 
\end{equation}
where $\mu_L$ and $\mu_k$ are, respectively, the chemical potentials of 
the left lead and the quantum dot when the dot contains $k$ electrons. 
The probability of the dot having $k$ electrons is denoted by $P_k$, and 
it is determined from a steady state rate equation.\cite{averin,whan} In 
Eq.~(\ref{eqn:IV}), the first term in the summation is the contribution 
to the current from the $k$-th electron on the dot which tunnels in from 
the left lead, given that the dot already has $k-1$ electrons on it. 
The second term corresponds to the reverse process. The net total current 
is the sum over all possible $k$ values, which depends on the applied voltage 
and the temperature. A more detailed account of our model and its 
application to various quantum-dot configurations are discussed 
elsewhere.\cite{whan} We now discuss results concerning 
the particular system of a quantum dot with superconducting leads. 

In Fig.~\ref{fig:IV-loT}, we show a typical low temperature 
current-voltage (I-V) characteristic of our system. Here the 
temperature $k_B T = 0.02 E_C^\ast$ ($E_C^\ast \equiv E_C + \epsilon$ 
is the spacing between chemical potential levels, and 
$E_C \equiv e^2/C_\Sigma $ is the charging energy), the superconducting 
energy gap  $2 \Delta = 0.3 E_C^\ast$ and the quantum energy level 
spacing in the dot $\epsilon = 0.2 E_C^\ast$. When the leads are 
superconducting (solid curve), the I-V curve consists of a series of 
sharp peaks spaced $\epsilon$ apart.\cite{note1} This is in contrast with 
the I-V curve of the same dot with normal metal leads (dashed curve), 
which has only gentle steps with the same spacing $\epsilon$.

The shape of these I-V curves and the various tunneling processes 
involved can be understood using the illustration shown in the inset, 
which is a sketch of the energy spectra of the dot and the leads. 
In this sketch, the narrow vertical lines represent the two tunneling 
barriers. The quasiparticle density of states for the superconducting 
leads are shown; the shaded region indicates occupied states and the 
unshaded, unoccupied states. The chemical potentials of the 
superconducting leads are shown as dotted lines in the center of the 
superconducting gap. The energy spectrum for the dot includes both 
the {\em excitation spectrum} (dotted horizontal lines) and the 
{\em addition spectrum} (heavy horizontal lines). The excitation spectrum 
has spacing $\epsilon$ and can be populated by, for example, thermal 
excitation of electrons within the dot without changing the electron 
number $N$. The addition spectrum, on the other hand, has spacing 
$ \mu_{N+1} - \mu_{N} = E_C^\ast$, since it includes the 
Coulomb charging energy that one has to overcome in order to $add$ an 
electron ($N \rightarrow N+1$) to the dot by means of tunneling. 
Assume at zero temperature and zero voltage that there are $N$ electrons 
on the dot and $\mu_L^0$ and $\mu_R^0$ are midway between $\mu_N$ 
and $\mu_{N+1}$. A dc voltage, $eV = \mu_L - \mu_R^0$ is applied to 
the left lead while the right lead remains at fixed voltage. In order 
for an electron to tunnel into the dot from the left lead, one needs 
to raise $\mu_L$, such that $\mu_L - \Delta \ge \mu_{N+1}$, or 
$V \ge V_t = {E_C^\ast}/2 + \Delta$. As soon as this threshold voltage, 
$V_t$, is reached, an electron on the occupied states of the left 
lead can tunnel into the dot at level $\mu_{N+1}$ and subsequently 
tunnel out to the right lead, resulting in a current flow which 
reflects the density of states of the left lead.\cite{note2} As the 
voltage is increased further, the first excited state of the 
($N+1$)-electron system (first dotted line above $\mu_{N+1}$) 
is reached. At this point, an electron can tunnel into the dot either 
through level $\mu_{N+1}$ or the first excited state above it, before 
tunneling out into the right lead. This accounts for the increased current 
at a voltage $\epsilon$ above the threshold voltage and having the shape 
again reflecting the density of states of the lead. We also see similar 
current rises corresponding to the second and higher excited states. 
However, as long as $\mu_L - \Delta < \mu_{N+2}$, the conduction cycle 
is always $N \rightarrow N+1 \rightarrow N$. When the voltage is high 
enough, such that $\mu_L - \Delta > \mu_{N+2}$ ($eV > eV_{t} + E_C^\ast$), 
one extra electron can tunnel into the dot (through $\mu_{N+2}$ or 
the excited states above it) before the ($N+1$)-th electron can tunnel out. 
Now the electron number on the dot is a random choice among $N$, $N+1$ and 
$N+2$ (with corresponding probabilities $P_{N}, P_{N+1}$ and $P_{N+2}$). 
This corresponds to a larger increase in the current at the voltage 
$V \approx 1.7$. As the voltage is increased further, $N$ will fluctuate 
among more and more possible configurations, and eventually the I-V will 
become essentially linear due to these fluctuations. 

The different current response with the superconducting and normal 
leads is simply the reflection of their respective density of states. 
The superconducting I-V has the sharp current rises and a higher 
threshold voltage because its density of state contains a gap and is 
singular at energies $|E| = \Delta$, both of which disappear in the 
normal leads where $\Delta = 0$. Quantum dots with discrete levels 
and normal leads have been studied previously by Averin {\it et al.}\cite{averin2}

The I-V curves become more complicated at higher temperatures due to 
thermal excitation of quasiparticles in the leads and electron 
population of the quantum dot excited states. Noting that there are 
many energy scales in our system and the fact that the superconducting 
energy gap decreases with increasing temperature,\cite{tinkham} we will 
concentrate in the regime, $E_C \gg k_B T \sim [\Delta (T), \epsilon]$. 
This parameter range is clearly relevant to the experiment in 
Ref.~\onlinecite{ralph}, where $E_C \approx 12$ meV, 
$\epsilon \approx 0.5$ meV, $\Delta (0) = 0.18$ and 
$k_B T_c \approx 0.1$ meV. 

In Fig.~\ref{fig:IV-hiT}(a) and (b), we show two I-V curves, both 
at relatively high temperature, such that in (a), $2 \Delta (T) > \epsilon$, 
while in (b), $2 \Delta (T) < \epsilon$. One of the intriguing 
features of these I-V curves is that in Fig.~\ref{fig:IV-hiT}(a) 
the narrow regions where the current is significantly suppressed 
{\em do not} correspond to the superconducting energy gap $2 \Delta (T)$. 
In fact, $2 \Delta (T)$ is much larger as indicated in the figure.          
In Fig.~\ref{fig:IV-hiT}(b), however, the relatively wide regions of 
current suppression do correspond to $2 \Delta (T)$. This seemingly 
contradictory phenomena has a simple explanation using the energy 
diagrams shown in the insets of Fig.~\ref{fig:IV-hiT}, which are 
the finite temperature versions of the diagram that we used 
previously in discussing Fig.~\ref{fig:IV-loT}. 

First of all, since $[k_B T/2\Delta (T), k_B T/ \epsilon] \sim 1$ (see 
the caption of Fig.~\ref{fig:IV-hiT}), we expect that there will be 
many quasiparticles that are thermally excited across the gap. Also, 
quite a few quantum levels in the dot near the chemical potential levels 
will be partially occupied due to thermal excitations within the dot. 
Therefore many quantum levels, both below and above the chemical potential 
positions of the dot, will become available for tunneling. However, due to 
conservation of energy, a level can contribute to the current 
only if there are electrons in the left lead that have the same 
energy. Since there are no electron states within the gap, 
whenever a quantum level in the dot lies inside the gap of the left 
lead, that level will be excluded from conduction. When we sweep the 
voltage, if $2 \Delta (T) > \epsilon$ as in Fig.~\ref{fig:IV-hiT}(a), there 
will always be either one or two quantum levels that fall inside the gap 
of the left lead. When the voltage is such that there are 
two levels lined up inside the gap and therefore excluded from 
conduction, the current will naturally drop in comparison with other 
voltage values where only one level is being excluded. This will explain 
Fig.~\ref{fig:IV-hiT}(a). The situation in Fig.~\ref{fig:IV-hiT}(b) is 
slightly different. Since we now have $2 \Delta (T) < \epsilon$, there 
can only be one or no level being excluded from conduction. When a level 
falls inside the gap, it has to traverse the entire gap region before 
it can join the tunneling process again. Thus the current suppression 
regions in Fig.~\ref{fig:IV-hiT}(b) have the same widths as the gap 
and are centered about every discrete level. 

Since the conduction in our system is carried out by tunneling processes,  
we expect there might be interesting ac properties to be explored when 
microwave radiation is coupled to the system. In particular, the 
photon-assisted tunneling phenomena, observed in large area 
superconductor-insulator-superconductor (SIS) tunnel 
junctions,\cite{dayem-martin,tien-gordon} and semiconductor quantum 
dots,\cite{kouwenhoven,blick} should also manifest itself in some 
fashion in the present system. We treat our system in the presence 
of microwave radiation using the Tien-Gordon formalism,\cite{tien-gordon} 
originally proposed to explain the photon-assisted tunneling phenomena 
observed in large area SIS junctions.\cite{dayem-martin} Recently, 
the Tien-Gordon theory has also been applied to photon-assisted 
tunneling in semiconductor quantum structures.\cite{kouwenhoven,hu}

We will treat the externally applied RF signal as an oscillating 
voltage, $V_{rf}\cos(2\pi \nu t)$, applied equally across each 
of the tunnel barriers. Then the Tien-Gordon theory leads to 
the following simple modification of the tunneling rates,\cite{kouwenhoven} 
\begin{equation}
\label{eqn:patGamma}
\Gamma (\mu_a - \mu_b , \alpha) = 
\sum_{n = -\infty}^{\infty} J_{n}^2(\alpha) \Gamma_{0} (\mu_a - \mu_b  + nh\nu)  
\end{equation}
where $\alpha = eV_{rf}/(h\nu)$, $J_{n}(x)$ are the Bessel functions 
of the first kind, and $\Gamma_{0}$ is the tunneling rate in 
the absence of RF radiation, given in Eq.~(\ref{eqn:golden-rule}). 

In Fig.~\ref{fig:pat}, we show I-V curves of our quantum dot with 
superconducting leads in the presence of RF radiation with frequency 
$h\nu/\epsilon = 2/3$ and two different power levels, along with the 
dark I-V ($\alpha = 0$). Due to the periodic spikes in the dark I-V, 
its photo-response is considerably more complicated than that of the 
conventional SIS tunnel junctions, whose dark I-V has a single 
discontinuous jump at $2\Delta$.\cite{tucker} For example, 
Fig.~\ref{fig:pat} illustrates that if the ratio $h\nu/\epsilon = p/q$ 
is a rational number, then the I-V, in sufficiently strong RF signal, 
will have periodic spikes with period $h\nu /p$, not $h\nu$. Even more 
complicated behavior is observed when $\epsilon$ and $h\nu$ are 
incommensurate.\cite{whan} We wish to point out that this effect, 
if confirmed by experiment, might find applications in millimeter 
wave detection and mixing schemes. In the weak signal limit, 
$\alpha \ll 1$, one expects a single peak below the dark I-V threshold 
voltage $V_t$, which will give information about the frequency as well 
as the amplitude of the RF signal. We found the sharpness of the current 
peaks in the superconducting-lead system makes it much more robust against 
thermal noise in comparison with its normal-lead counterpart. 
  
In summary, we have carried out numerical simulations of single electron 
transport through a quantum dot with superconducting leads. Our 
current-voltage characteristics calculated at low temperature 
is in good qualitative agreement with the experimental observation of 
Ralph {\it et al.}\cite{ralph} In addition our analysis show that at 
higher temperatures, thermal excitation of quasiparticles 
in the leads and thermal population of the excited quantum levels within the 
quantum dot should lead to interesting changes in the I-V curves. We also 
predict that when RF radiation is coupled to the system, the photon-assisted 
tunneling phenomena should manifest itself by producing extra periodic 
structures in the I-V curves, which might be useful in the millimeter 
wave detector/mixer applications. Due to the presence of many different 
characteristic energy scales, the rich dynamical properties of this 
system demands more exploration.

\acknowledgments
We acknowledge helpful discussions with Dan Ralph, Chuck Black and 
David Carter. This project is supported by the National Science 
Foundation under grant DMR-9402020 and the U.S. Air Force Office 
of Scientific Research under grant F49620-95-1-0311.

\begin{figure}
\caption{Low temperature I-V characteristics of a quantum dot with 
superconducting leads (solid curve) and normal metal leads 
(dashed curve). The temperature is $k_B T = 0.02 E_C^\ast$, and the 
superconducting energy gap in the leads is $2\Delta = 0.3 E_C^\ast$. The 
quantum level spacing is, $\epsilon = 0.2 E_C^\ast$. The inset is a sketch 
of the energy spectra in the leads and the dot. Note the quantum dot 
energy spectrum includes the excitation spectrum (with spacing $\epsilon$), 
and addition spectrum (with spacing $E_C^\ast = e^2/C + \epsilon$).
\label{fig:IV-loT}} 
\end{figure}

\begin{figure}
\caption{I-V characteristics at higher temperatures for (a) 
$2\Delta (T) > \epsilon$ and (b) $2\Delta (T) < \epsilon$.  
The relevant energy ratios are: (a) $T/T_c = 0.815$, 
$k_B T/E_C^\ast = 0.07$, $k_B T/2\Delta(T) = 0.311$, 
$k_B T/\epsilon = 0.35$, $2\Delta (T)/\epsilon = 1.14 $; 
(b) $T/T_c = 0.931$, $k_B T/E_C^\ast = 0.08$, 
$k_B T/2\Delta(T) = 0.576$, $k_B T/\epsilon = 0.40$, 
$2\Delta (T)/\epsilon = 0.70 $. The vertical dotted lines are 
drawn at the positions of the quantum levels ($\epsilon$ apart).
\label{fig:IV-hiT}}
\end{figure}
 
\begin{figure}
\caption{Photon assisted tunneling in quantum dots with superconducting 
leads. The dark I-V (solid curve) has the same parameter values as 
Fig.~\ref{fig:IV-loT}(a). The dotted curve and the dashed curve, 
respectively correspond to power levels, $\alpha = eV_{rf}/h\nu = 2.5$, 
and $\alpha = 0.5$, with $h\nu /\epsilon = 2/3$ for both curves. 
\label{fig:pat}} 
\end{figure}


\begin{references}
\bibitem{note0}Strictly speaking, using Fermi-Dirac distribution to 
describe a quantum dot is an approximation[Ref.~\onlinecite{beenakker}]. 
Nevertheless, our experiences indicate that it is a reasonably good 
approximation, especially for dots that contain large numbers of 
electrons.
\bibitem{note1}For convenience, we attributed all of $C_\Sigma$ to the self-capacitance of the dot, but in a tunnel junction system, $C_\Sigma = C_L + C_R$. In order to compare the I-V curves in this paper with those measured in 
an experiment, one needs to multiply all the voltage scales with the 
``lever arm'' factor, $\eta = C_\Sigma /C_R$, if the voltage on the left 
lead is being swept.
\bibitem{note2}The gap structure and the entire density of states on the 
right lead have little effect in the I-V curve when it is held at constant 
voltage. This is because when an electron tunnels into the dot from the left 
lead, its energy always corresponds to empty states that are much higher 
than the gap in the right lead.
\end{references}
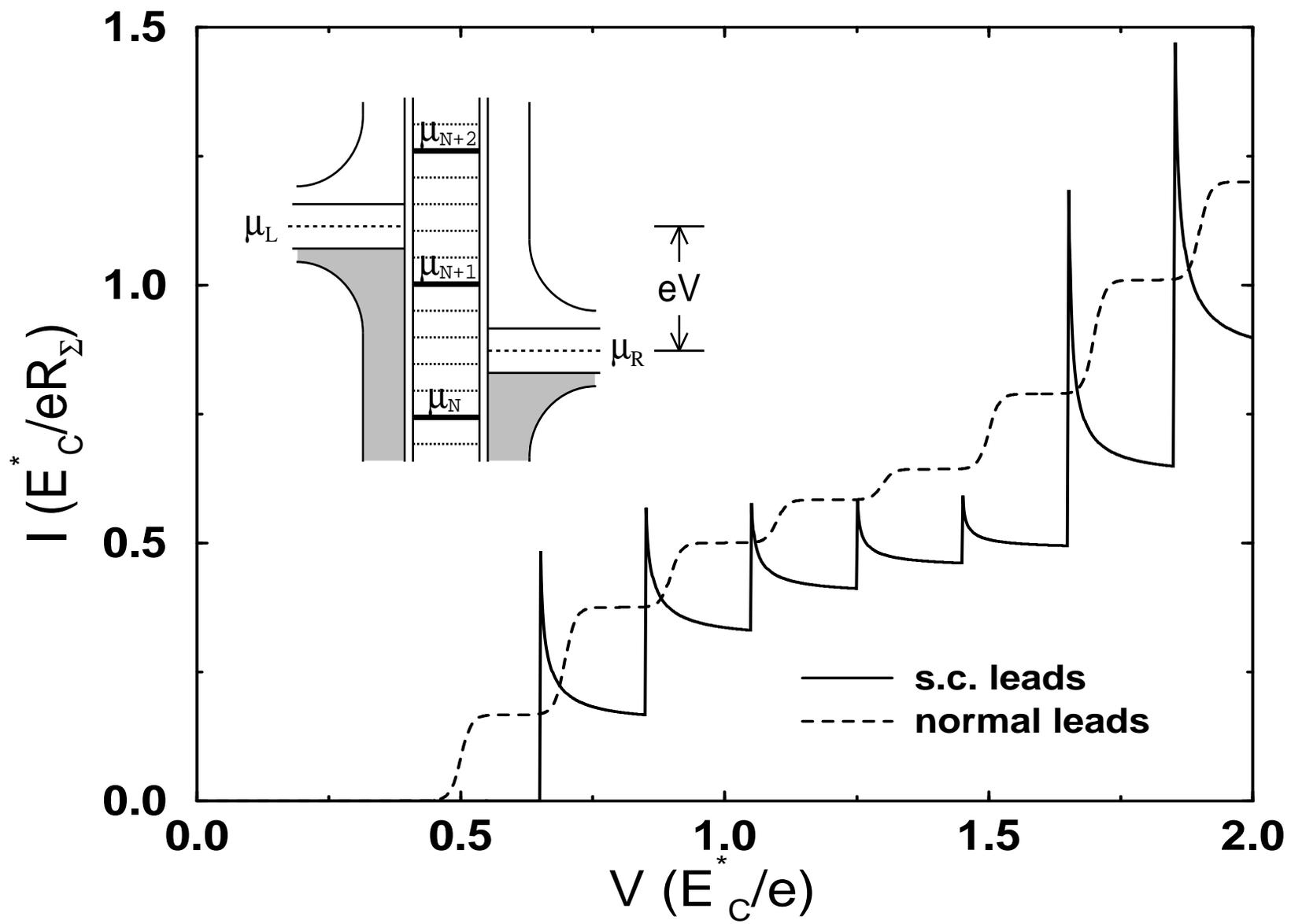
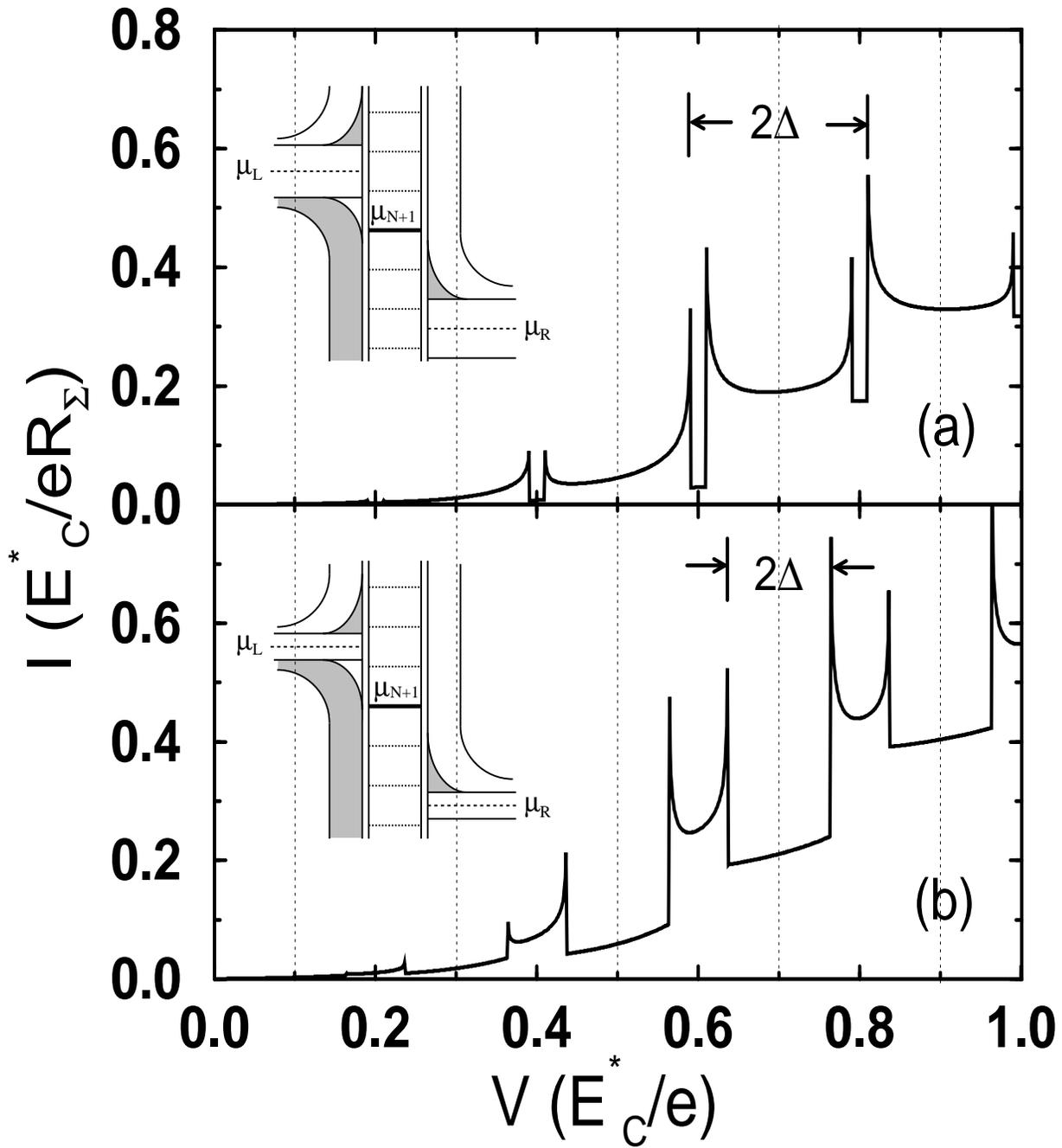
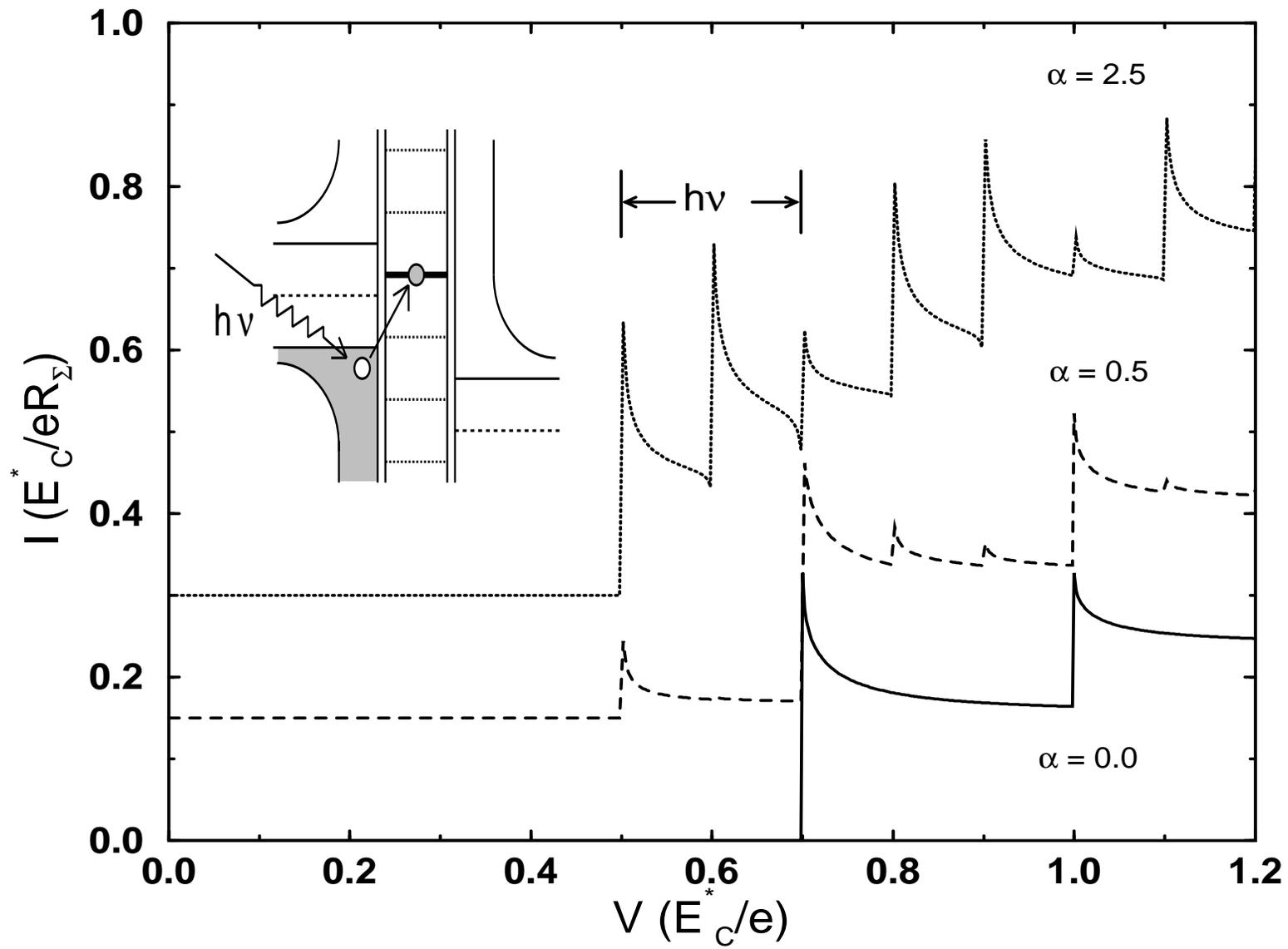
\end{document}